\documentclass{article}

\usepackage{arxiv}

\usepackage[utf8]{inputenc} 
\usepackage[T1]{fontenc}    
\usepackage{hyperref}       
\usepackage{url}            
\usepackage{booktabs}       
\usepackage{amsfonts}       
\usepackage{nicefrac}       
\usepackage{microtype}      
\usepackage{xcolor}
\usepackage{graphicx}
\usepackage{subcaption}
\usepackage{multicol}

\usepackage{listings} 

\newenvironment{Figure}
  {\par\medskip\noindent\minipage{\linewidth}}
  {\endminipage\par\medskip}

\title{CARlasso: An R package for the estimation of sparse microbial networks with predictors}

\author{
Yunyi Shen \\
Department of Statistics \\
Department of Forest and Wildlife Ecology\\
University of Wisconsin-Madison
\And 
Claudia Sol\'is-Lemus\thanks{Corresponding author: solislemus@wisc.edu}\\
Wisconsin Institute for Discovery \\
Department of Plant Pathology \\
University of Wisconsin-Madison
}
\newcommand{\runningtitle}{\texttt{CARlasso} R Package}

\begin{document}
\maketitle

\begin{abstract}
\textbf{Summary:} 
Microbiome data analyses require statistical tools that can simultaneously decode microbes' reactions to the environment and interactions among microbes.
We introduce \texttt{CARlasso}, the first user-friendly open-source and publicly available R package to fit a chain graph model for the inference of sparse microbial networks that represent both interactions among nodes and effects of a set of predictors. Unlike in standard regression approaches, the edges represent the correct conditional structure among responses and predictors that allows the incorporation of prior knowledge from controlled experiments. 
In addition, \texttt{CARlasso} 1) enforces sparsity in the network via LASSO; 2) allows for an adaptive extension to include different shrinkage to different edges; 3) is computationally inexpensive through an efficient Gibbs sampling algorithm so it can equally handle small and big data; 4) allows for continuous, binary, counting and compositional responses via proper hierarchical structure, and 5) has a similar syntax to \texttt{lm} for ease of use. The package also supports Bayesian graphical LASSO and several of its hierarchical models as well as lower level one-step sampling functions of the CAR-LASSO model for users to extend. 
 

\textbf{Availability and Implementation:} The R package \texttt{CARlasso} is a publicly available and open-source under the GNU General Public License (version 3). Installation can be done via CRAN or the GitHub repository directly (\url{https://github.com/YunyiShen/CAR-LASSO}).

\textbf{Contact:} yshen99@wisc.edu or solislemus@wisc.edu


\end{abstract}


\begin{multicols}{2}
\section{Introduction}

Understanding the composition of microbial communities and what environmental or experimental factors play a role in shaping this composition is crucial to comprehend biological processes in humans, soil and plants alike.
Mathematically, we represent the microbial community as a sparse network where nodes represent microbial taxa and edges represent some form of correlation. We also want to include a second type of nodes for the environmental or experimental predictors that we believe are shaping the microbial community. For example, on soil microbiome data, the predictors could be related to the presence of certain fertilizer, weather or humidity in the soil \cite{Allsup2019, Rioux2019, Lankau2020, Lankau2020b}.

Currently, users would need to fit a multiresponse regression model and treat it as a graphical model to obtain a network with nodes for responses and predictors. In \cite{shen2021bayesian}, we argued that this approach is flawed given that the regression coefficient matrix does not encode the conditional dependence structure between response and predictor nodes.

Here, we introduce \texttt{CARlasso}, an R package implementing the chain graph model in \cite{shen2021bayesian} that allows users to infer a complex network structure that represents both interactions among microbial taxa and effects of a set of predictors. 
Our package is the first software implementation that allows users to 
estimate a network that represents the conditional dependence structure of a multivariate response (e.g. abundances of microbes) while simultaneously estimating the conditional effect of a set of predictors that influence the network (e.g. diet, weather, experimental treatments).

In addition to providing a more sensible representation compared to standard multiresponse linear regression, our software has six main advantages: 1) it guarantees a sparse solution via Bayesian LASSO; 2) it allows users to choose an adaptive extension to use different shrinkage on different edges; 3) it allows for continuous, binary, count and compositional responses; 4) it is scalable for big data given its \texttt{C++} backend implementation and efficient Gibbs sampler; 5) it is easy to use as \texttt{lme4} \cite{lme4} providing a straight-forward transition for those using the multiresponse regression; 6) it provides a visualization function which allows an easy and attractive representation of the network.

\section{Implementation and general features}

\noindent \textbf{Input data.}
Just like standard regression approaches, the input data for our method is a data frame with rows corresponding to samples and columns corresponding to covariates. We have two sets of covariates: responses (denoted $y_1,y_2,\dots,y_k$) and predictors (denoted $x_1,x_2,\dots,x_p$). For microbiome studies, the responses are compositional in nature as they usually refer to relative abundances of microbial species and the predictors are any environmental or experimental factors that are expected to alter the microbial communities. Our method allows for continuous, binary and counting responses as well.

\noindent \textbf{Methods.} The main function of the \texttt{CARlasso} package is \texttt{CARlasso}$(y_1+\cdots+y_k \sim x_1+\cdots+x_p$, data, link, adaptive$)$ where \textit{data} corresponds to the input dataframe, \textit{link} corresponds to the link function with "identity" for normal response (default), "probit" for binary, "log" for counting, "logit" for compositional. Note that for "logit", the last response is used as reference. Finally, \textit{adaptive} is a Boolean option and if it is set to true, then the shrinkage will be specific to every edge (as opposed to the same shrinkage for every edge in the network). By default, we set adaptive to false. In addition, we present a similar function to perform Bayesian graphical LASSO \texttt{bGlasso} with the similar syntax and two one-step sampling functions (\texttt{rCARlasso\_} and \texttt{rCARAlasso\_}) for non-adaptive and adaptive version of CAR-LASSO for advanced users to build their own hierarchical models.

\noindent \textbf{Results.} The \texttt{CARlasso} function produces a \texttt{carlasso\_out} object which contains, among many elements, point estimates for the posterior mean of the precision matrix ($\Omega$) and the posterior mean of the regression coefficient matrix ($\mathbf{B}$). We implemented a plot function that will read a \texttt{carlasso\_out} object and will produce the sparse network with nodes for responses (circles) and nodes for predictors (triangles). The edges are color-coded with red edges corresponding to positive correlations and blue edges corresponding to the negative correlations. The strength of the edge effect is represented by the width of the edge.

\section{Application to human gut microbiome}

We incorporate in the package a sample data related to the study of gut microbiota in the Irish elderly \cite{claesson2012gut}. This study is based on pyrosequencing of 16S rDNA amplicons from faecal samples collected from 178 elderly Irish citizens and 13 healthy young control subjects. Other environmental factors were recorded such as diet, BMI and age.

We fit the \texttt{CARlasso} model to this sample data and plot the result (Figure \ref{fig:gut}) with the following commands:
\begin{lstlisting}[language=python]
res <- CARlasso(Alistipes + Bacteroides +
    Eubacterium + Parabacteroides +
    all_others ~ BMI + Age +
    Gender + Stratum,
    data = mgp154, link = "logit", 
    adaptive = TRUE, n_iter = 5000, 
    n_burn_in = 1000, thin_by = 10)

plot(res)
\end{lstlisting}

\begin{Figure}
    \centering
    \includegraphics[scale=0.25]{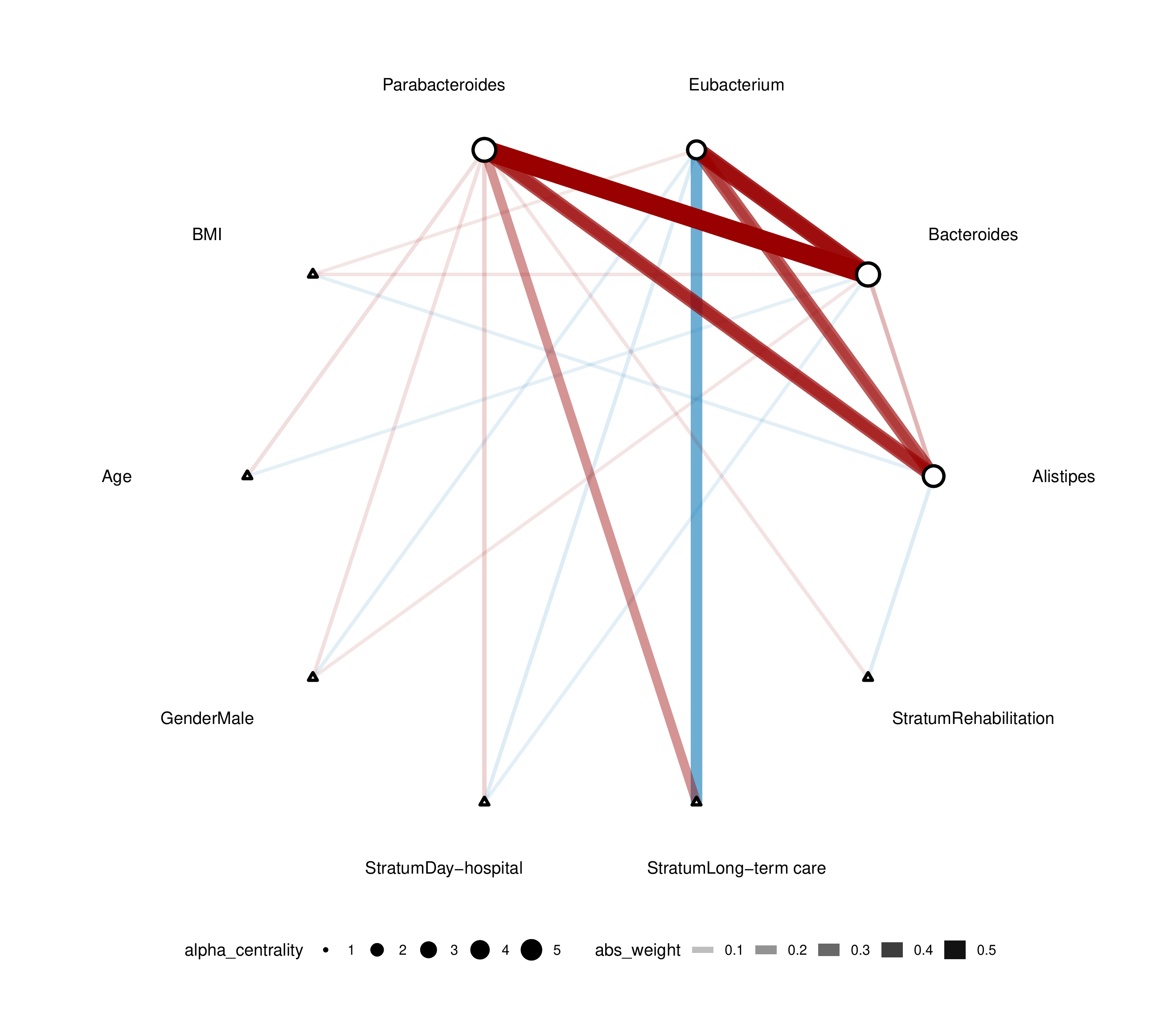}
    \captionof{figure}{Reconstructed genus network for human gut microbiota. Triangle nodes correspond to predictors and circle nodes correspond to relative abundances of genus. The node size on the circle nodes correspond to the $\alpha-$centrality values \cite{bonacich2001eigenvector}. The width of the edges correspond to the absolute weight, and the color to the type of interaction (red positive, blue negative).}
    \label{fig:gut}
\end{Figure}

\section{Conclusions}

We have developed \texttt{CARlasso}, an R implementation of the chain graph model in \cite{shen2021bayesian} that allows users to estimate a microbial sparse network with nodes for the responses (e.g. abundances of microbes) and nodes for predictors shaping the microbial community (e.g. weather, diet). 
By utilizing a similar syntax to \texttt{lme4} \cite{lme4}, users can easily run our methods on their respective microbiome data. In addition, our visualization tools allows for a clear representation of the microbial community and the effect of each predictor in the model.

The package is available on the Comprehensive R Archive Network
and on GitHub (\url{https://github.com/YunyiShen/CAR-LASSO}).

\section*{Acknowledgements}  
\noindent \textit{Financial Support:} This material is based upon work support by the National Institute of Food and Agriculture, United States Department of Agriculture, Hatch project 1023699.
This work was also supported by the Department of Energy [DE-SC0021016 to C.S.L.].

\vspace{0.25cm}
\noindent \textit{Conflict of Interest:} None declared.


\bibliographystyle{unsrt}  
\bibliography{refs}  
\end{multicols}


\end{document}